\documentclass[12pt]{jpconf}

\usepackage[OT2,T1]{fontenc}
\usepackage[english]{babel}
\usepackage[latin1]{inputenc}
\usepackage[charter]{mathdesign}
\usepackage[scaled=0.9]{helvet}
\usepackage{url,color}
\usepackage{enumerate}
\usepackage{verbatim}  

\usepackage{longtable}
\usepackage{tabularx}
\usepackage{multicol} 
\usepackage{multirow}
\usepackage{booktabs} 
\usepackage{ltxtable}
\usepackage{hhline}
\usepackage{bm}

\usepackage[final]{graphicx}
\usepackage{epsf}
\usepackage{wrapfig}
\usepackage{subfigure}
\usepackage{float}

\usepackage{caption}
\usepackage{mathrsfs}  

\usepackage[numbers,sort&compress]{natbib}

\usepackage{icomma}                        
\usepackage[pdfpagelabels,plainpages=false]{hyperref}  

\definecolor{uni}{rgb}{0.490,0.604,0.667}
\definecolor{ippblue}{cmyk}{1,0.45,0.04,0.}

\hypersetup{pdftitle={Double-resonant fast particle-wave interaction},
            pdfauthor={Mirjam Schneller},
            pdfsubject={Double-resonant fast particle-wave interaction},
            pdfcreator={\LaTeX2e},
colorlinks, linkcolor=black, urlcolor=blue, citecolor=green}

\usepackage{fancybox}

\begin{document}

\newcommand{\qp}{$q$ profile}
\newcommand{\bef}{$\beta_{\mathrm{f}}$}

\title[Nonlinear alfv{\'e}nic fast particle transport and losses]{Nonlinear alfv{\'e}nic fast particle transport and losses}

\author{M Schneller$^1$, Ph Lauber$^1$, M Garc{\'i}a-Mu{\~n}oz$^1$, M Br{\"u}dgam$^1$  and S G{\"u}nter$^1$}
\address{$^1$ Max Planck Institute for Plasma Physics, EURATOM Association, Boltzmannstr.\ 2, 85748 Garching, Germany}
\ead{\mailto{mirjam.schneller@ipp.mpg.de}}

\begin{abstract}
Magnetohydrodynamic  instabilities like Toroidal Alfv{\'e}n Eigenmodes or core-localized modes such as 
Beta Induced Alfv{\'e}n Eigenmodes and Reversed Shear Alfv{\'e}n Eigenmodes 
driven by fast particles can lead to significant redistribution and losses in fusion devices. 
This is observed in many ASDEX Upgrade discharges. 
The present work aims to understand the underlying resonance mechanisms, especially in 
the presence of multiple modes with different frequencies. Resonant mode coupling mechanisms 
are investigated using the drift kinetic \textsc{Hagis} code [Pinches 1998].\\
Simulations were performed for different plasma equilibria, in particular for different $q$ profiles, 
employing the availability of improved experimental data. A study was carried out, 
investigating double-resonant mode coupling with respect to various overlapping scenarios. 
It was found  that, depending
on the radial mode distance, double-resonance is able to enhance growth rates as well as 
mode amplitudes significantly. 
Small radial mode distances, however can also lead to strong nonlinear mode stabilization of 
a linear dominant mode.\\
With the extended version of \textsc{Hagis}, losses were simulated and 
directly compared with experimental loss measurements. The losses' phase 
space distribution as well as their ejection signal is consistent with experimental data. Furthermore,
it allowed to characterize them as prompt, resonant or stochastic. It was found that especially 
in multiple mode scenarios (with different mode frequencies), abundant incoherent losses occur 
in the lower energy range, due to a broad phase-space stochastization. 
The incoherent higher energetic losses are ``prompt'', i.e.\ their initial energy is too large for 
confined orbits.
\end{abstract}

\section{Introduction}
    Fusion devices contain fast particle populations due to external plasma
    heating and (eventually) fusion-borne $\alpha$-particles.
    Fast particle populations can interact with global electromagnetic waves, 
    leading to the growth of MHD-like and kinetic instabilities -- e.g., Toroidicity
    Induced Eigenmodes (TAE) \cite{Cheng85}, Reversed Shear 
    Alfv{\'e}n Eigenmodes (RSAE) \cite{Berk01}
    or Beta Induced Alfv{\'e}n Eigenmodes (BAE) \cite{Heidbrink93}.
    In this work, drift-kinetic fast particle simulations performed
    with the \textsc{Hagis} code \cite{Pinches98} are 
    carried out to obtain a deeper understanding of the dynamics of wave-fast particle interaction in
    scenarios with two modes of different frequencies: what is the mode coupling mechanism, and 
    how does it dependent on radial mode distance?
    In a second study, the extended version of \textsc{Hagis} \cite{mwb_phd} is used to 
    simulate fast particle losses and 
    to compare them with experimental loss measurements \cite{garcia10} in ASDEX Upgrade (AUG). 
    The losses' phase space distribution as well as their ejection signal allows to characterize them as 
    prompt, resonant or stochastic.

\section{Simulation Tool and Experimental Reference Case}
    The numerical investigations are performed with 
    the \textsc{Hagis} Code \cite{Pinches98},
    a nonlinear, drift-kinetic, perturbative Particle-in-Cell code, that models the interaction between
    a distribution of energetic particles and a set
    of Alfv{\'e}n Eigenmodes.
    The plasma equilibrium for \textsc{Hagis} is based 
    on the \textsc{Cliste} \cite{Carthy12} and \textsc{Helena} \cite{Huysmans91} codes.
    The data for the MHD equilibria originate from the
    ICRH minority heated ASDEX Upgrade discharge \#23824, at times  $t=1.16$\ s 
    and $t=1.51$\ s. At the earlier time point, the \qp\ is slightly inverted ($q_0 = -1.55$, $q_{\mathrm{min}}=-1.43$)
    at the later time point, it is monotonic, 
    with lower absolute values ($q_0 =
    q_{\mathrm{min}}=-0.97$)\footnote{Note that the \qp\ is negative here, due
      to the AUG current direction.}. This particular reference scenarios were chosen due
    to the availability of detailed experimental data concerning fast particle-wave interaction and losses.
    The comparison between these data and numerical results is subject of \sref{numloss}.
 
\section{Numerical Study on Double-Resonance}
   \paragraph{Theoretical Picture}
    Theory (e.g.\ Ref.\ \cite{Berk92}) predicts that conversion of free energy
    to wave energy is enhanced in a multiple-mode scenario, i.e., the interaction of multiple modes produces
    energy conversion rates higher than that which would be achieved with each mode acting independently. 
    This effect is called ``double-resonance'' and can be partially explained by the principle of 
    gradient driven mode growth -- according to\footnote{$s$ refers to the radial coordinate as 
    the square root of the normalized poloidal flux: 
    $s=(\psi_{\mathrm{pol}}/\psi_{\mathrm{pol, edge}})^{1/2}$.}:
    $\gamma \propto \nabla f(s)$ \cite{Fu89} -- which can be extended to 
    multiple modes \cite{Berk90, Berk92}.    
    This picture of \emph{\textbf{gradient driven double-resonance}} is based on the
    precondition that modes share resonances in the same phase space area.
    Through the resulting redistribution by each mode, a steeper gradient
    is produced at the other mode's position, enhancing its drive. The radial 
    overlapping of modes leads then to a much larger conversion of free energy to wave
    energy.\\
    However, this mechanism can only work if the modes not only share resonance regions in
    phase space, but if there is also \emph{spatial} mode overlap in 
    the radial direction. In Ref.\ \cite{mwb_phd} simulations were carried
    out, finding a double-resonant effect also without this precondition.
    Furthermore, a superimposed oscillation on the modes' amplitudes was observed,
    clearly indicating mode-mode interaction. The modes without radial overlap
    are then coupled radially through the particles' trajectories: A population
    of particles that shares resonances in phase space with both modes and passes 
    both modes' location at once, can transfer energy from one mode to the other \cite{mwb_phd}.
    Due to the particle orbits' width, it is not necessary that
    both modes have a radial overlap. In the following, this mechanism is called 
    \emph{\textbf{inter-mode energy transfer}}.

   \paragraph{Simulation Conditions}
    As the first question, the understanding of double mode resonance is very fundamental, 
    the simulations were performed under quite simple physical conditions:
    radial particle distribution $f(\psi)$ 
    with constant gradient (to avoid different mode drive at different radial mode positions
    only due to a steeper gradient) and a slowing down function as energy distribution function.
    The volume averaged fast particle beta is \bef=1\%.
    The particles are distributed isotropically in pitch angle (as 
    e.g., fusion borne $\alpha$-particles would be).
     \begin{figure}[H]
        \centering
        \subfigure[analytical perturbation]{\includegraphics[width=0.45\textwidth,height=0.13\textheight]{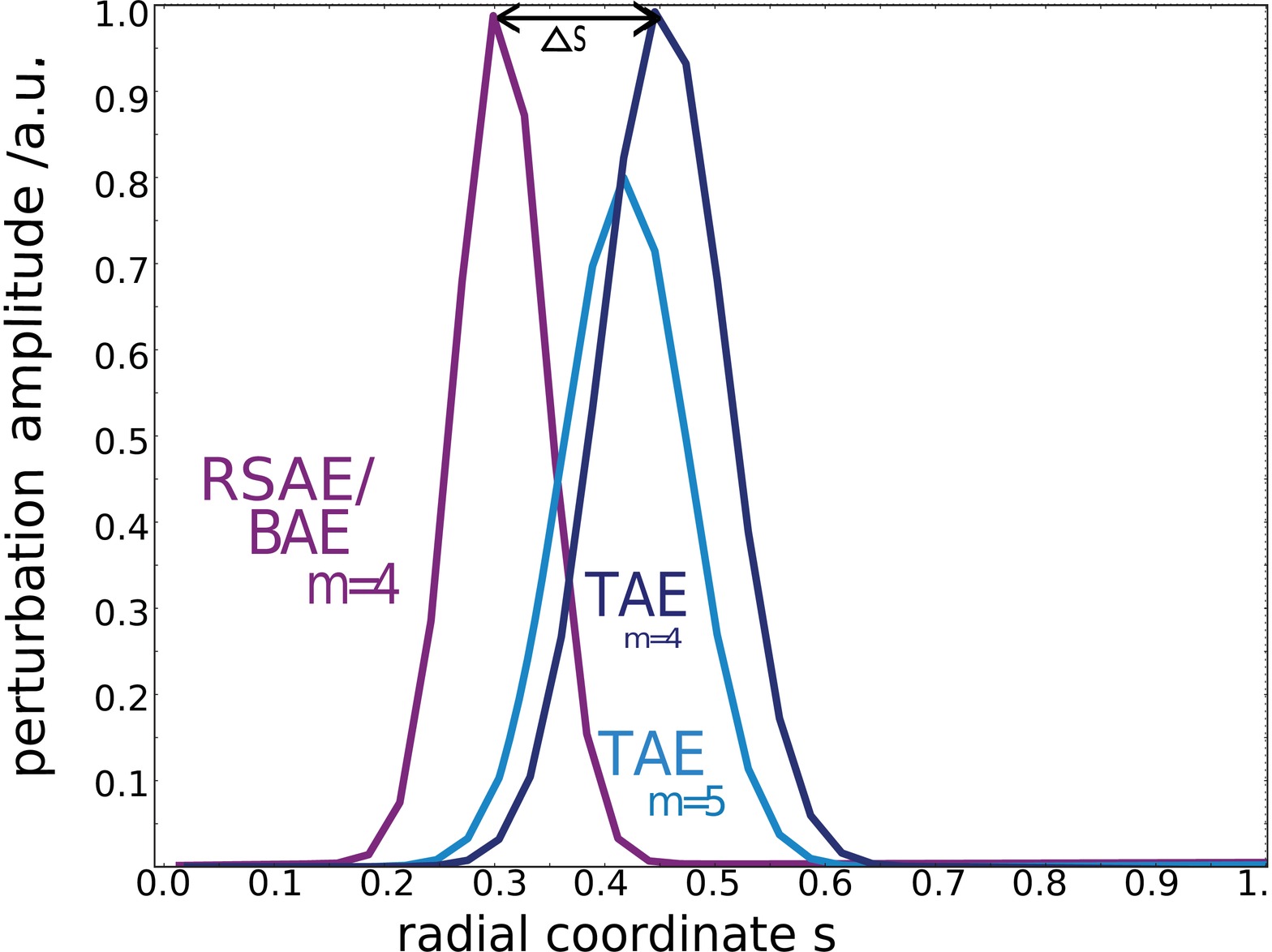}}
        \hfill\subfigure[numerical perturbation]{\includegraphics[width=0.45\textwidth,height=0.13\textheight]{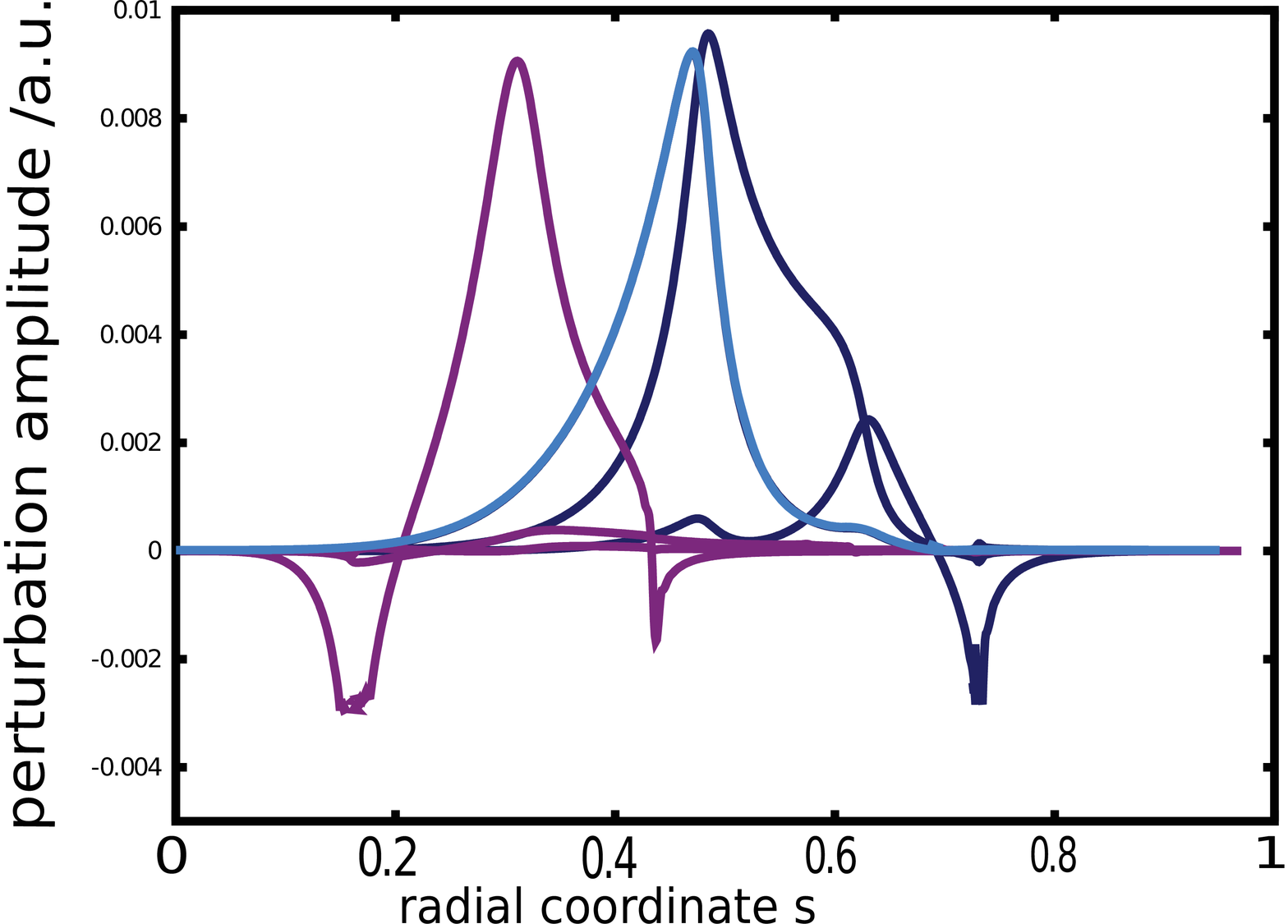}}
        \caption{\itshape Two Alfv{\'e}nic modes: (a) gauss-shaped perturbation as used in the following simulations, 
          (b) for comparison: perturbations occurring in AUG discharge \#23824, as calculated numerically with 
          \textsc{Ligka} \cite{Lauber07}.}
        \label{pert_exam}
     \end{figure}
    As MHD perturbations, analytic, gauss shaped functions are used (see \fref{pert_exam}), 
    without background damping. The mode frequencies are chosen to
    match experimental data: one high frequency mode $120$\ kHz (TAE) 
    and one lower frequency mode at $55$\ kHz (possible RSAE or BAE).

\begin{figure}[H]
      \centering
      \subfigure[\itshape Growth rates enhancement]{\includegraphics[width=0.47\textwidth,height=5.5cm]{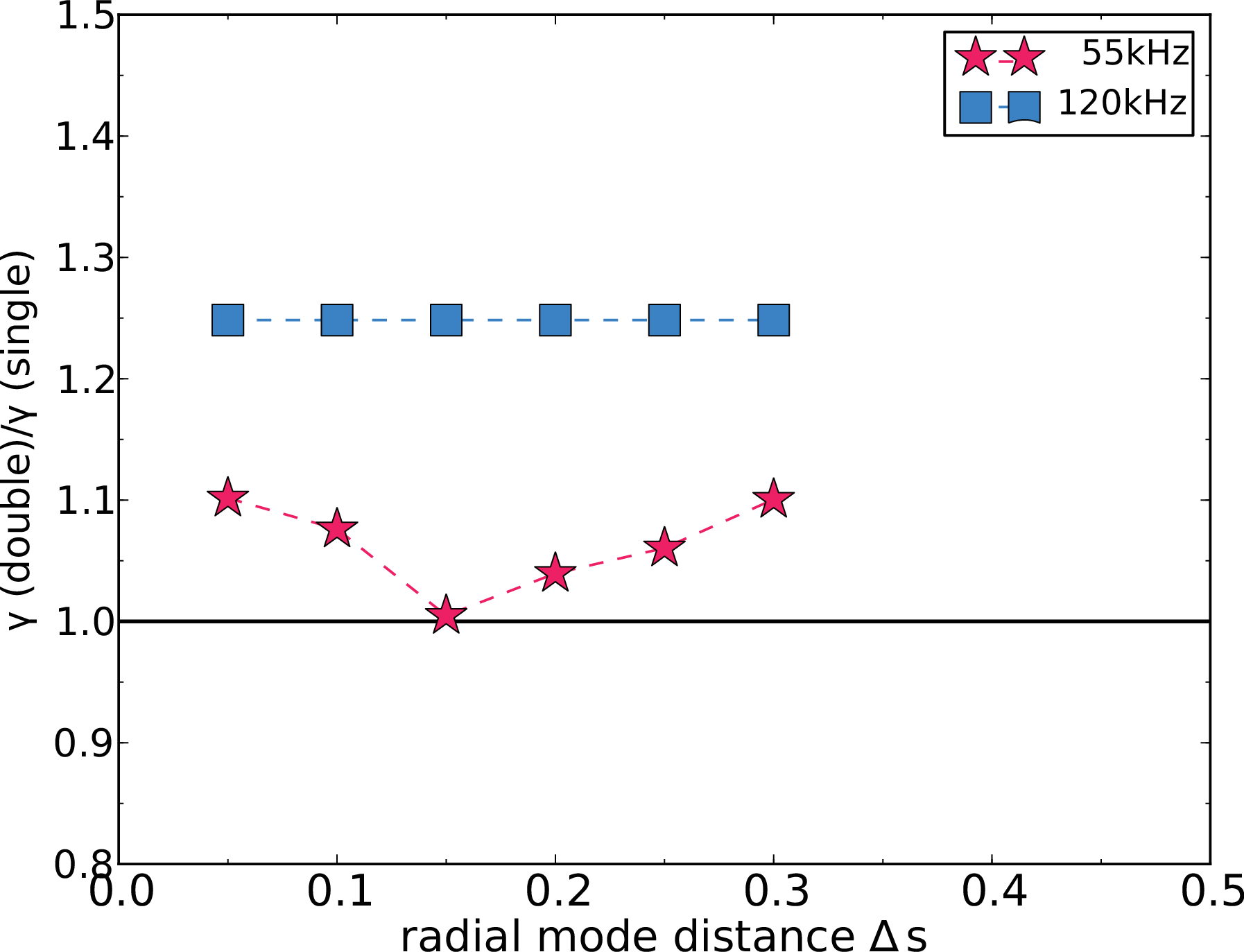}}
      \hspace{0.7cm}\subfigure[\itshape Amplitudes enhancement (at $t\approx 300$ T)]{\includegraphics[width=0.47\textwidth,height=5.5cm]{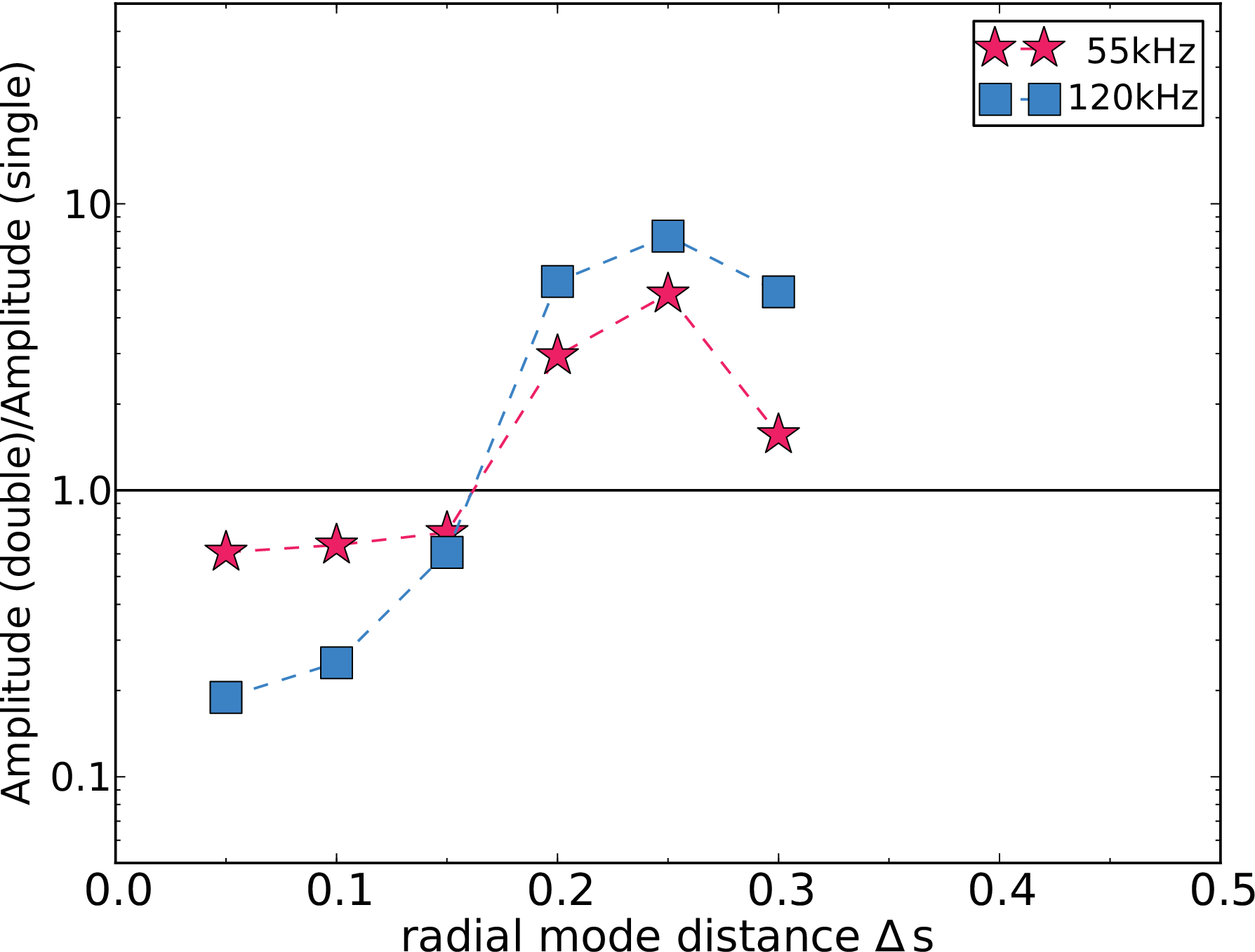}}
      \caption{\itshape Mode evolution scanned over the radial mode distance $\Delta s$ in the
        inverted \qp\ case. Depicted is the $\Delta s$ dependence of the ratios 
        of growth rates (a) and amplitudes (b) in double mode 
        simulations over those from single mode simulations. Red star: RSAE, blue square: TAE.
        One can clearly see that larger radial mode distances lead to higher amplitudes,
        whereas amplitudes are even lower than in the single mode case for $\Delta s \le 0.15$.
        The linear growth rates, however, are higher than in the single mode simulation throughout the
        $\Delta s$ range. The RSAE growth rate experiences a small drop at $\Delta s \approx 0.15$.}
      \label{deltas-scan}
    \end{figure}

    \paragraph{Simulation Results}
    A scan over the radial mode distance reveals an effective double-resonance as shown in 
    \fref{deltas-scan}: depicted are the ratios of the linear growth rates (a) and the amplitudes 
    (b) in the double mode case vs.\ the single 
    mode case over the radial mode distance $\Delta s$. 
    The amplitude level was compared after $\approx 300$ TAE periods ($=~2.5$\ ms) of 
    simulation time. This time is sufficient for the single amplitudes to saturate,
    but still significantly below energy slowing down time.
    One can see that the \textbf{growth rates} of
    both the TAE and the RSAE are enhanced in all double mode cases compared to the single mode ones.
    However, the growth rate of the outer TAE is enhanced most strongly and independently 
    of the radial 
    mode distance -- i.e.\ gradient driven double-resonance works even if there is no radial mode overlap.  
    In contrast, the enhancement of the inner and weaker RSAE decreases with the radial 
    mode distance for small $\Delta s$. Then it increases again for $\Delta s > 0.15$. 
    These larger mode distances match the double-resonant
    particle orbits and therefore enable inter-mode energy exchange, driving the weak mode. For
    higher $\Delta s$, the larger, i.e.\ higher energetic orbits fit the mode distance
    and lead to even more energy exchange.
    Furthermore, with larger radial mode distances, the modes are able to tap energy from a wider 
    gradient region. The \textbf{amplitude} ratios, however, are even lowered in the double-resonant case 
    compared to the respective single mode levels, if the radial mode distance is small. This happens 
    due to the mutual gradient depletion at the 
    other mode's radial position. If the modes are at a larger radial distance ($\Delta s > 0.15$), the
    double mode scenario amplitudes are much higher compared to the single mode amplitudes, both for the
    TAE and the RSAE. Both modes drive each other -- most for a radial distance of about 
    $\Delta s \approx 0.25$.
    It is important to note that the distance $\Delta s=0.25$ giving maximum amplitude ratios depends
    strongly on the absolute mode positions with respect to the radial distribution function, 
    and especially on the amplitude regime (stochastic or non-stochastic) of each mode. 
    The same applies for the value $\Delta s=0.15$ at which the transition towards
    double-resonant amplitude enhancement takes place.\\\\
    \paragraph{Summary (I)} Growth rates are
    generally enhanced by the presence of another mode, whereas for the amplitudes, this is only
    the case, if the modes have sufficiently large radial distance. For small distances, modes at radial 
    positions, where the initial distribution function is already relatively flat, double-resonance 
    leads to strong mode stabilization with respect to the single mode case. 
    If the amplitudes are enhanced, their amplification level is, 
    however, mainly determined by  whether the mode reaches  the stochastic regime. 
    The stochastic threshold is reached much earlier or even only if a second mode is present.

\section{Fast Particle Losses in Numerical Simulations vs.\ Experimental Measurements}\label{numloss}

    \paragraph{Experimental Loss Observations}
    In the AUG discharge \#23824, during the inverted \qp\ equilibrium at $t=1.16$\ s, 
    a high loss signal is observed at the Fast Ion Loss Detector (FILD), with
    the majority of losses characterized as \emph{incoherent}. Later, at $t=1.51$\ s, during the 
    monotonic \qp\ equilibrium, only very few losses occur, and all are identified as \emph{coherent} \cite{garcia10}.
    In the following, simulations within this experimental reference frame are presented, 
    that were performed to identify the origin of the different loss types.
       
\begin{figure}[H]
      \centering
      \subfigure[\itshape FFT of losses' time signal during resonant phase]{\includegraphics[width=0.9\textwidth, height=2.8cm]{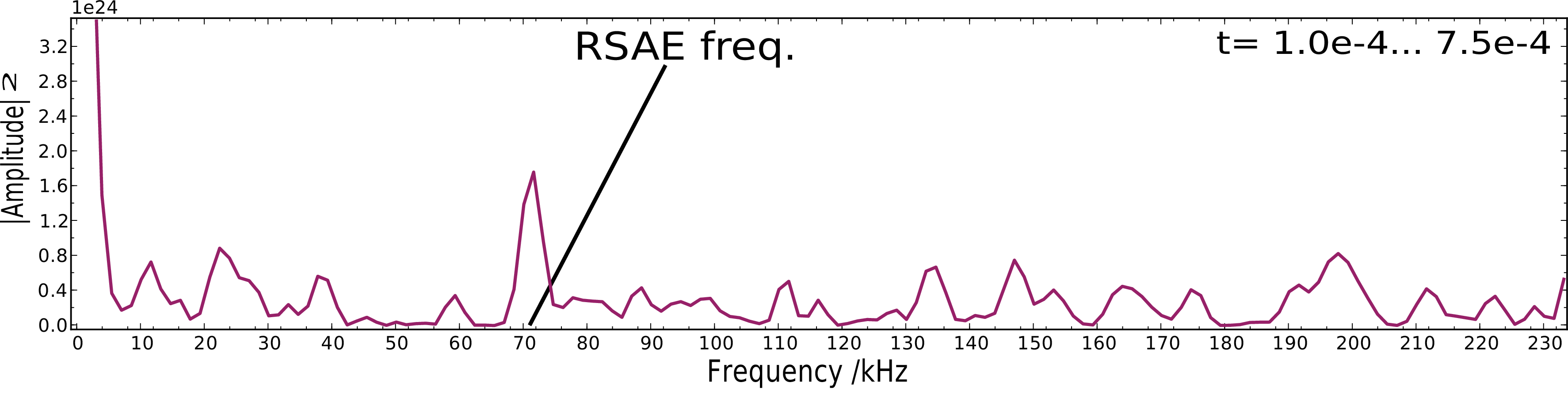}}\\
      \vspace{0.7cm}
      \subfigure[\itshape FFT of losses' time signal during stochastic phase]{\includegraphics[width=0.9\textwidth, height=3.2cm]{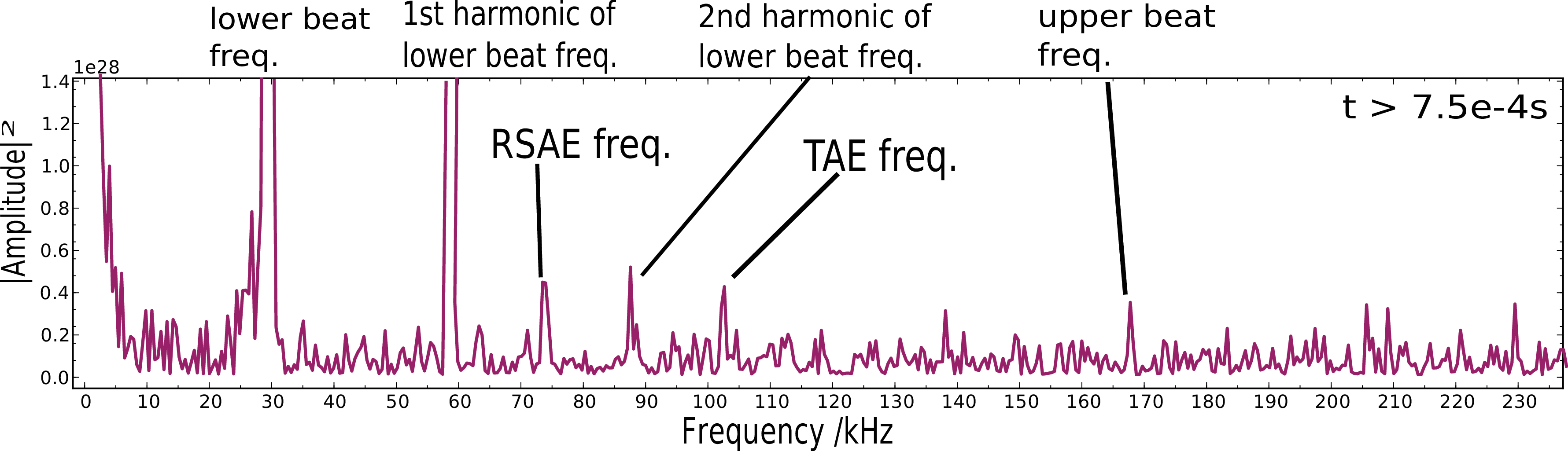}}
      \caption{\itshape Fourier spectra of losses time signal. (a) in the resonant phase:
        a significant number of particles is ejected with the frequency of the dominating RSAE (70\ kHz).
       (b) in the stochastic phase: the beat frequency (30\ kHz)
       is by far dominating (and its higher harmonics).
       All frequencies experience a slight down-shift.}
      \label{run0343_tloss-fft}
    \end{figure}

   \paragraph{Simulation Conditions}
    The simulations presented in this section are carried out with an extended \textsc{Hagis} version \cite{mwb_phd},
    including the vacuum region, and are based on more realistic plasma conditions:
    The radial particle loading follows a Fermi-like potential law $f(s) = (1-s^2)^5$,
    that decays to zero at $s=0.6$. It was chosen in a way that
    does not produce prompt losses in the simulations with monotonic
    \qp, as seen in the experiment.
    Furthermore, the markers are initialized in pitch and poloidal 
    angle distribution according to an analytical function that models the distribution created by ICRH
    (main heating method in AUG \#23824). \bef\ is set to $0.1\%$.
    As MHD perturbation, the analytic, gauss functions (\fref{pert_exam}) are used, 
    without background damping. The mode frequencies, positions and widths are chosen to
    be consistent with experimental data: at $t=1.16$\ s, a 120\ kHz TAE at $s=0.45$ and a 55\ kHz RSAE 
    at $s=0.35$, whereas at $t=1.51$\ s, there is a TAE of 180\ kHz at $s=0.6$ and
    a BAE of 80\ kHz at $s=0.35$ \cite{Lauber09}.
    These data were obtained from Fourier spectrograms of the SXR signal, 
    Mirnov Coils and the FILD signal \cite{garcia11}.
   \begin{figure}[H]
      \centering
      \subfigure[\itshape redistribution in phase space]{\includegraphics[width=0.46\textwidth, height=6cm]{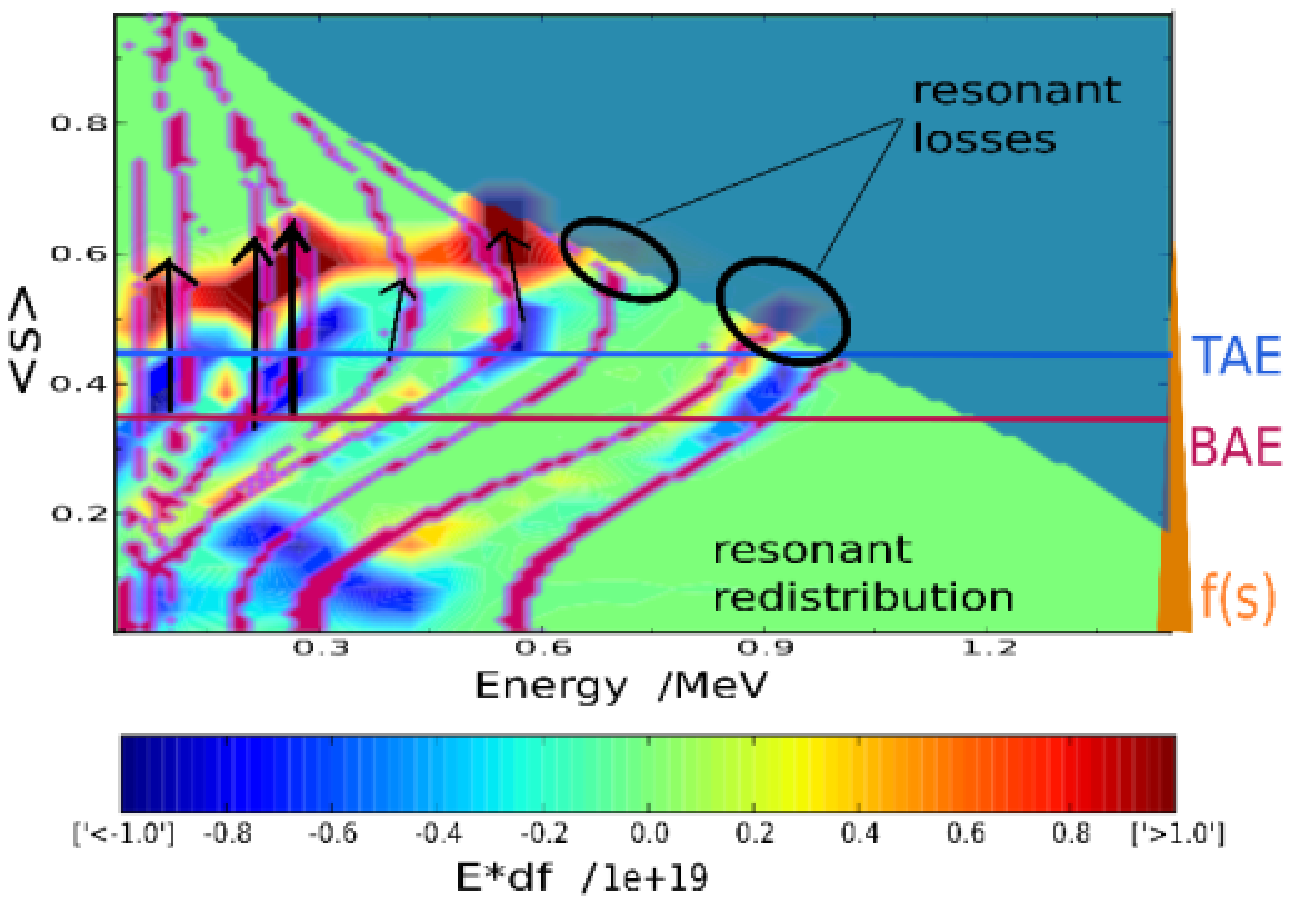}}
      \hspace{0.5cm}\subfigure[\itshape losses in phase space]{\includegraphics[width=0.44\textwidth, height=6cm]{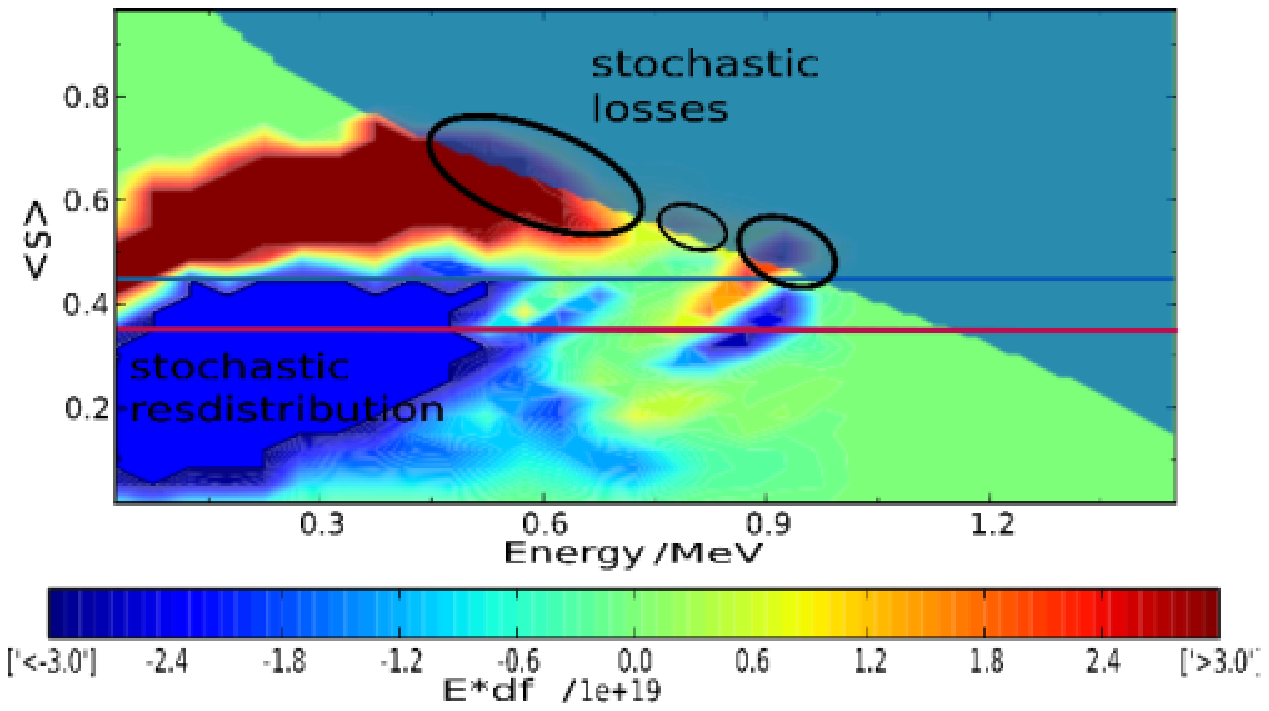}}
      \caption{\itshape Double mode simulation in the inverted \qp\ equilibrium: 
        Redistribution in $E$-$s$ space during (a): redistribution phase ($t\approx 0.6$\ ms),
        (b) stochastic phase ($t\approx 1.8$\ ms)
        Red: particles accumulate, blue: particles move away. One can see the good accordance of redistribution areas
        (black arrows) and the resonance lines (pink) in the modes' vicinity (horizontal lines).
        The redistribution in the areas, where the resonance lines meet the loss boundary (black circles)
        coincide with the losses (see \fref{run0255+0256_tEloss}a).}
      \label{run0256_Esdf-2115+resplot+lossres}
    \end{figure}
    \paragraph{Simulation Results}
    Consistent with results from the numerical study, it is found that the mode amplitudes grow slower 
    with the monotonic \qp\ and do not reach the stochastic
    threshold (at $\approx 2\cdot 10^{-3} \delta $B/B). 
    A superimposed oscillation frequency is visible in both scenarios, originating from the double-resonance mechanism
    of inter-mode energy exchange. Its frequency matches with the beat 
    frequency of both modes.
    In the inverted \qp, the modes reach the stochastic regime. There, particle redistribution 
    in phase space takes place broadly over the whole energy range (\fref{run0256_Esdf-2115+resplot+lossres}b) at 
    the modes' radial position. Slightly before, during the resonant phase, however, redistribution in phase
    space occurs along the resonance lines (\fref{run0256_Esdf-2115+resplot+lossres}a).
    The energy of the losses obtained during this ``redistribution phase'' (see \fref{run0255+0256_tEloss}a) matches roughly 
    with the energy, at which the higher resonance lines meet the loss boundary. This indicates, 
    that losses in this time period are \emph{resonant}, i.e.\ \emph{coherent} losses. In the simulation with the monotonic \qp,
    there is no redistribution into the loss region. Thus, very few losses appear, consistent with experimental measurements.\\
    To learn more about the identity of the \emph{incoherent}
    losses, their \textbf{phase space pattern} is analyzed in the following.
    The distribution function of the fast particles is not known exactly,
    therefore, the markers are loaded according to an estimated distribution
    function. To learn how sensitive the 
    loss pattern is with respect to the parameters defining the distribution function,
    a scan is performed over the radial, as well as over the
    poloidal distribution function.
     Combining the results of both the radial and the poloidal marker
     loading scan, it turns out that there is a phase space region
     that cannot be covered by prompt losses in neither loading scenario:
     the phase space region of energies below 600\ keV (45\ mm gyroradius) at 
     negative pitch angles smaller than 63\textdegree. These losses are
     very likely to be stochastic losses. For
     the rest of the phase space, it is not possible to discern prompt from
     later losses yet.
    \begin{minipage}{1\textwidth}
    \begin{minipage}{0.5\textwidth}
    \begin{figure}[H]
      \centering
      \includegraphics[width=1\textwidth]{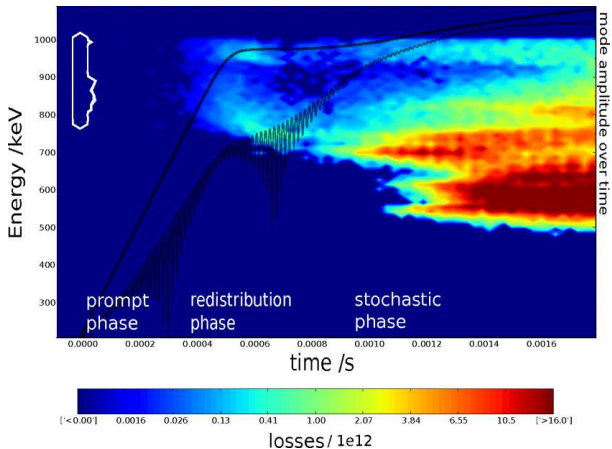}
      \caption{\itshape Lost particles at the first wall over simulation time in the inverted \qp:
         in the stochastic phase, losses spread over a large energy range, even to lower energies around 500\ keV. 
         For comparison, the edge of the phase space region of the prompt losses in this 
         simulation is shown (white lines). Black lines: amplitude evolution of RSAE (upper) and TAE (lower).}
      \label{run0255+0256_tEloss}
    \end{figure}
    \end{minipage}
    \vspace{0.4cm}
    \begin{minipage}{0.5\textwidth}
    \begin{figure}[H]
      \centering
      \includegraphics[width=0.8\textwidth]{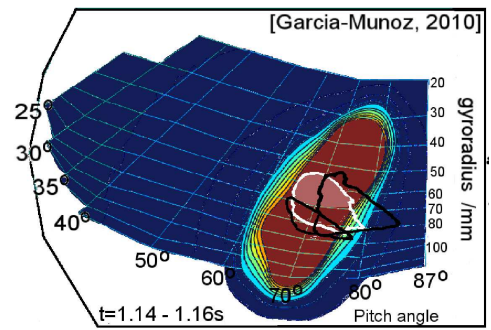}
      \caption{\itshape Fast ion losses in phase space (AUG discharge \#23824):
        colors give the FILD measured loss pattern \cite{garcia10}, lines
        give the boundary of the numerical calculated loss region.
        Black lines: upper estimate for the (drift corrected) 
        border of the prompt losses appearance resulting from the 
        radial and poloidal marker loading scan. White lines: same for 
        the losses during the stochastic phase of the MHD perturbations.
        The lighter shaded area indicates the peak region of stochastic losses.}
      \label{run0294+0302_num+exp}
     \end{figure}
     \end{minipage}
    \end{minipage}
     \Fref{run0294+0302_num+exp} shows the good accordance of the
     numerically calculated losses and the experimentally measured ones: 
     the whole loss area as well 
     as the peak region of the numerical values lies almost
     entirely within the respective experimentally measured region of phase
     space.
     The experimental peak being slightly broader results mainly from
     over exposure of the FILD loss plate. The higher energetic losses
     (i.e.\ the pattern at gyroradii > 90\ mm or energy > 1\ MeV) do not appear in the simulation,
     as the fast particle energy was cut at 1\ MeV. 
     The lower energetic losses below 50\ mm gyroradius are expected to
     appear when simulating with more than two modes at lower frequencies (work in progress).\\
    When looking closely at the loss time traces, one can perceive an ejection modulation. 
    A \textbf{Fourier analysis of the loss signal} gives a peak at 65\ kHz, which is exactly the beat frequency. 
    In the above simulation, the mode frequencies result
    in beat frequencies of 65\ kHz and 175\ kHz. Unfortunately, the RSAE frequency is very close to the 
    lower beat frequency, and the TAE frequency almost coincides with the first harmonic of it.
    Therefore, another simulation was performed, using $\omega_{\mathrm{RSAE}} = 70$ and 
    $\omega_{\mathrm{TAE}} = 100$\ kHz.  
    A Fourier analysis of the loss signal is performed 
    for the resonant phase (\fref{run0343_tloss-fft}a),
    and the stochastic phase (\fref{run0343_tloss-fft}b).
    In the resonant phase, the RSAE frequency peak at around 70\ keV is visible, whereas the TAE's 
    amplitude is still too small to produce significant losses. 
    Later on in the simulation, the TAE grows towards the same amplitude range as the 
    RSAE, likewise, the beat frequency peak becomes visible in 
    the Fourier spectrum. This is consistent with FILD spectra, showing the beat frequency as well.
    \paragraph{Summary (II)}
    With the extended version of \textsc{Hagis}, losses were simulated and 
    compared with experimentally observed losses at the Fast Ion Loss Detector.
    The simulated losses' phase space pattern coincides very well with the 
    experimental one. Especially in multi-mode
    scenarios with different mode frequencies, stochastic redistribution sets in over a broad
    energy range, leading to lower energetic losses, that are incoherent. The resonant losses
    appear from the late linear phase on, mainly in the high energy regime, showing 
    good coherence with the modes' frequencies and especially their beat frequencies.
    The higher energetic part of experimentally measured incoherent losses was identified
    as mainly prompt losses. The internal redistribution, as well as the losses can be 
    understood very well as processes in phase space, when combining resonance lines, 
    loss boundary and radial positions of the Alfv{\'e}nic modes.\\\\
\bibliographystyle{iopart-num}
\bibliography{literature}

\providecommand{\newblock}{}
\begin{thebibliography}{10}
\expandafter\ifx\csname url\endcsname\relax
  \def\url#1{{\tt #1}}\fi
\expandafter\ifx\csname urlprefix\endcsname\relax\def\urlprefix{URL }\fi
\providecommand{\eprint}[2][]{\url{#2}}

\bibitem{Cheng85}
Cheng C~Z, Chen L and Chance M 1985 {\em Ann. Phys.\/} {\bf 161} 21 -- 47

\bibitem{Berk01}
Berk H~L, Borba D~N, Breizman B~N, Pinches S~D and Sharapov S~E 2001 {\em Phys.
  Rev. Lett.\/} {\bf 87}

\bibitem{Heidbrink93}
Heidbrink W~W, Strait E~J, Chu M~S and Turnbull A~D 1993 {\em Phys. Rev.
  Lett.\/} {\bf 71}(6) 855--858

\bibitem{Pinches98}
Pinches S, Appel L, Candy J, Sharapov S, Berk H, Borba D, Breizman B, Hender T,
  Hopcraft K, Huysmans G and Kerner W 1998 {\em Comput. Phys. Commun.\/} {\bf
  111} 133 -- 149

\bibitem{mwb_phd}
Br{\"u}dgam M 2010 Ph.D. thesis Technische Universit{\"a}t M{\"u}nchen

\bibitem{garcia10}
{Garc{\'i}a-Mu{\~n}oz M et al} 2010 {\em Phys. Rev. Lett.\/} {\bf 104}(18)

\bibitem{Carthy12}
{Mc Carthy P J} and {the ASDEX Upgrade Team} 2012 {\em Plasma Phys. Control.
  Fusion\/} {\bf 54} 015010

\bibitem{Huysmans91}
{Huysmans G T A et al} 1990 Conf. on Computational Physics Proc. World
  Scientific, Singapore (AIP)

\bibitem{Berk92}
Berk H~L, Breizman B and Ye H 1992 {\em Phys. Rev. Lett.\/} {\bf 68}(24)
  3563--3566

\bibitem{Fu89}
Fu G~Y and Dam J~W~V 1989 {\em Phys. Fluids B\/} {\bf 1} 1949--1952

\bibitem{Berk90}
Berk H~L and Breizman B~N 1990 {\em Phys. Fluids B\/} {\bf 2} 2226--2252

\bibitem{Lauber07}
{Lauber Ph}, G{\"u}nter S, K{\"o}nies A and Pinches S~D 2007 {\em J. Comp.
  Phys.\/} {\bf 226} 447 -- 465

\bibitem{Lauber09}
{Lauber Ph et al} 2009 {\em Plasma Phys. Control. Fusion\/} {\bf 51} 124009

\bibitem{garcia11}
{Garc{\'i}a-Mu{\~n}oz M et al} 2011 {\em Nuclear Fusion\/} {\bf 51}

\end{thebibliography}

\end{document}